\documentclass[a4paper,superscriptaddress,aps,prl,twocolumn,floatfix,citeautoscript,showpacs,floatfix]{revtex4-1}
\usepackage{graphicx}
\usepackage{epsfig}
\usepackage{latexsym}
\usepackage{amsmath,bm}

\usepackage{natbib}
\usepackage{booktabs}  % nice tables
\usepackage{dcolumn}
\newcommand{\etal}{\textit{et al.~}}

\newcommand{\GW}{\ensuremath{GW}}

\newcommand{\be}{\begin{equation}}
\newcommand{\ee}{\end{equation}}
\newcommand{\bea}{\begin{eqnarray}}
\newcommand{\eea}{\end{eqnarray}}
\newcommand{\ba}{\begin{array}}
\newcommand{\ea}{\end{array}}

\def\vec#1{\mathchoice{\mbox{\boldmath$\displaystyle#1$}}
{\mbox{\boldmath$\textstyle#1$}}
{\mbox{\boldmath$\scriptstyle#1$}}
{\mbox{\boldmath$\scriptscriptstyle#1$}}}

\newcommand{\D}{{\rm d}}

\newcommand{\br}{\vec{r}}

\renewcommand {\vec} [1] {{\bm #1}}

\begin{document}

\title{Density-based mixing parameter for hybrid functionals}

\author{Miguel A. L. Marques}
\affiliation{LPMCN, Universit\'e Claude Bernard Lyon I and CNRS, 69622
Villeurbanne, France}
\affiliation{European Theoretical Spectroscopy Facility (ETSF)}

\author{Julien Vidal}
\affiliation{LPMCN, Universit\'e Claude Bernard Lyon I and CNRS, 69622
Villeurbanne, France}
\affiliation{Laboratoire des Solides Irradi\'es, \'Ecole Polytechnique, CNRS,
CEA-DSM, 91128 Palaiseau, France}
\affiliation{European Theoretical Spectroscopy Facility (ETSF)}
% \affiliation{Institute for Research and Development of Photovoltaic Energy
% (IRDEP), UMR 7174 CNRS/EDF/ENSCP, 6 quai Watier, 78401 Chatou, France}

\author{Micael Oliveira}
\affiliation{LPMCN, Universit\'e Claude Bernard Lyon I and CNRS, 69622
Villeurbanne, France}
\affiliation{European Theoretical Spectroscopy Facility (ETSF)}

\author{Lucia Reining}
\affiliation{Laboratoire des Solides Irradi\'es, \'Ecole Polytechnique, CNRS,
CEA-DSM, 91128 Palaiseau, France}
\affiliation{European Theoretical Spectroscopy Facility (ETSF)}

\author{Silvana Botti}
\affiliation{Laboratoire des Solides Irradi\'es, \'Ecole Polytechnique, CNRS,
CEA-DSM, 91128 Palaiseau, France}
\affiliation{LPMCN, Universit\'e Claude Bernard Lyon I and CNRS, 69622
Villeurbanne, France}
\affiliation{European Theoretical Spectroscopy Facility (ETSF)}
% \date{\today}

\begin{abstract}
  A very popular ab-initio scheme to calculate electronic properties
  in solids is the use of hybrid functionals in density functional
  theory (DFT) that mixes a portion of Fock exchange with DFT
  functionals.  In spite of their success, a major problem still
  remains, related to the use of one single mixing parameter for all
  materials. Guided by physical arguments that connect the mixing
  parameter to the dielectric properties of the solid, and ultimately
  to its band gap, we propose a method to calculate this parameter
  from the electronic density alone. This method is able to cut
  significantly the error of traditional hybrid functionals for large
  and small gap materials, while retaining a good description of
  structural properties.  Moreover, its implementation is simple and
  leads to a negligible increase of the computational time.
\end{abstract}


\maketitle

Density functional theory (DFT) is one of the major achievements of
theoretical physics in the last decades. It is now routinely used to
interpret experiments or to predict properties of novel materials.
The success of DFT relies on the
Kohn-Sham (KS) scheme and the existence of good approximations for the
unknown exchange and correlation (xc) functional.  In the standard KS
formulation the xc potential is local and static. Since the original
suggestion of the local-density approximation (LDA)~\cite{kohn65}, a
swarm of functionals has been proposed in the
literature~\cite{scuseria05}.  In the ab-initio study of solids, the
Perdew, Burke and Ernzerhof~\cite{perdew96} (PBE) parametrization of
the xc functional has been for many years the default choice for many
applications. A good functional must yield ground states properties
(like structural parameters), while it is expected that the KS gap and
true quasiparticle gap differ by the derivative
discontinuity~\cite{delta_xc}.  Indeed, for semiconductors and
insulators PBE yields good structural properties and KS band-gap
energies that are at best half of their experimental value.  
To obtain both the ground state and quasiparticle energies
correctly within one and the same formalism, one can resort to, e.g., 
a many-body \GW\ calculation~\cite{hedin65,aulbur00}.  However, $GW$ is by all
measures an expensive technique, with a very unfavorable scaling with
the number of atoms in the unit cell. It is therefore unpractical for
the study of band structures of large systems and clearly
prohibitive regarding total energy calculations even for simple
realistic systems.

Much of the computational effort in \GW\ comes from the dynamically
screened Coulomb interaction $W$. It has therefore been crucial to
explore to which extent dynamical effects are mandatory, or whether
non-locality is the dominating characteristic. The move from local KS
potential to non-local functionals has first been pushed forward in
Quantum Chemistry, where today the so-called hybrid functionals are
very popular. These functionals mix a fraction $\alpha$ of Fock
exchange with a combination of LDA and generalized gradient (GGA)
functionals. The application of hybrid functionals to the solid state
had a much slower start~\cite{ernzerhof99,brothers08}.  The situation
changed recently, helped by the wider availability of computer codes
that support hybrids~\cite{vasp} and the continued increase of
computational power covering the additional cost with respect to a
local potential. Besides yielding good structural
properties~\cite{paier07} hybrids have proved to correct to a large
extent the band-gap problem~\cite{brothers08,paier08}.  Another
landmark came with the introduction of screened hybrids~\cite{heyd03}. These
functionals lead to faster calculations and improved band-gaps, 
especially for small band-gap systems. Furthermore, by screening
the Coulomb interaction at large distances, they also give access to
metals.

The intuition lying behind an hybrid functional is rather clear.
While LDA or GGA calculations strongly underestimate the gap,
Hartree-Fock calculations overestimates it typically by more than a
factor of two. By changing the mixing from 0 to 1, one has a
continuous change between local KS and Hartree-Fock, and an
essentially linear variation between the respective gaps.  Therefore,
to obtain the correct experimental gap one simply has to use an
appropriate mixing in the functional. This value can be determined
from a fit to a series of systems, and set to around $\alpha\sim
0.2-0.3$. This choice gives very good results for a large class of
systems, but it usually fails when the gap is very large or very
small. But what is the physical meaning of the mixing parameter? To
answer this question, it can be instructive to move away from a
generalized KS picture~\cite{seidl96}, and consider the hybrid as an
approximation to the self-energy $\Sigma$ . In the GW approximation
the latter can be written as
\begin{equation}
  \label{eq:qp}
  \Sigma(\br,\br';\omega) = \Sigma_\text{sX}(\br,\br') + 
  \Sigma_\text{rest}(\br,\br';\omega)
  \,,
\end{equation}
with $\Sigma_\text{sX}(\br,\br')$ being the statically
screened-exchange (sX) term, and $\Sigma_\text{rest}(\br,\br';\omega)$
containing the static Coulomb hole and dynamical contributions. If the
screening in the sX term is replaced by an effective static dielectric
constant $\epsilon_\infty = 1/\alpha$, and $\Sigma_\text{rest}$ is
modeled by the static and local part of the hybrid
functional~\cite{gygi_baldereschi} the quasi-particle equation has the
same form as the generalized KS equation solved for hybrid
functionals~\cite{note}. From these arguments we can conclude that the
physical value for the mixing parameter $\alpha$ is related to the
inverse of the dielectric constant of the material at hand. Such a
link has been suggested to explain also variations of band gaps with
respect to small structural changes~\cite{vidal10}.

\begin{figure}
\begin{center}
 \includegraphics[width=.5\textwidth]{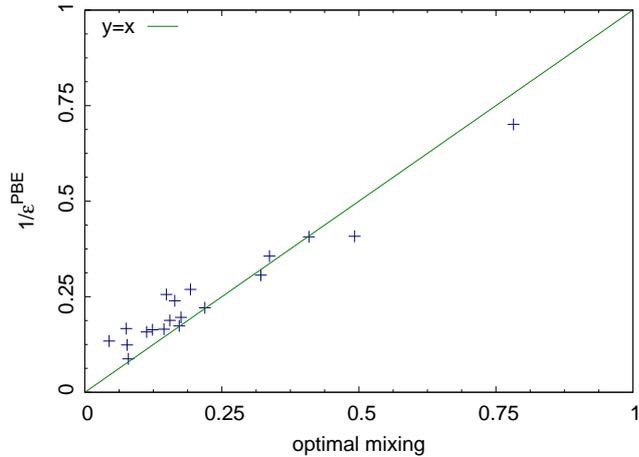}
 \caption{(Color online) 
   Inverse of the dielectric constant calculated
   with the PBE as a function of the optimal mixing
   parameter. The line $\alpha_\text{opt}=1 /
   \epsilon^\text{PBE}_\infty$ is a guide to the eye.
 \label{fig:epsvsmix}
}
\end{center}
\end{figure}

In Fig.~\ref{fig:epsvsmix} we show that, for a large range of
different materials, the value of $1/\epsilon_\infty$ -- obtained with
{\sc abinit}~\cite{abinit} within the PBE approximation -- is
approximately proportional to the optimal mixing parameter
$\alpha_\text{opt}$. The latter, calculated with
the computer code {\sc vasp}~\cite{vasp}, is defined as the fraction $\alpha$ of
Fock exchange of a PBE0~\cite{ernzerhof99} hybrid functional that
reproduces the experimental band-gap of the material.   The correlation is evident,
despite the fact that the DFT-PBE calculations systematically overestimate the
dielectric constants with respect to experiment.

\begin{figure*}
\begin{center}
 \includegraphics[width=.45\textwidth]{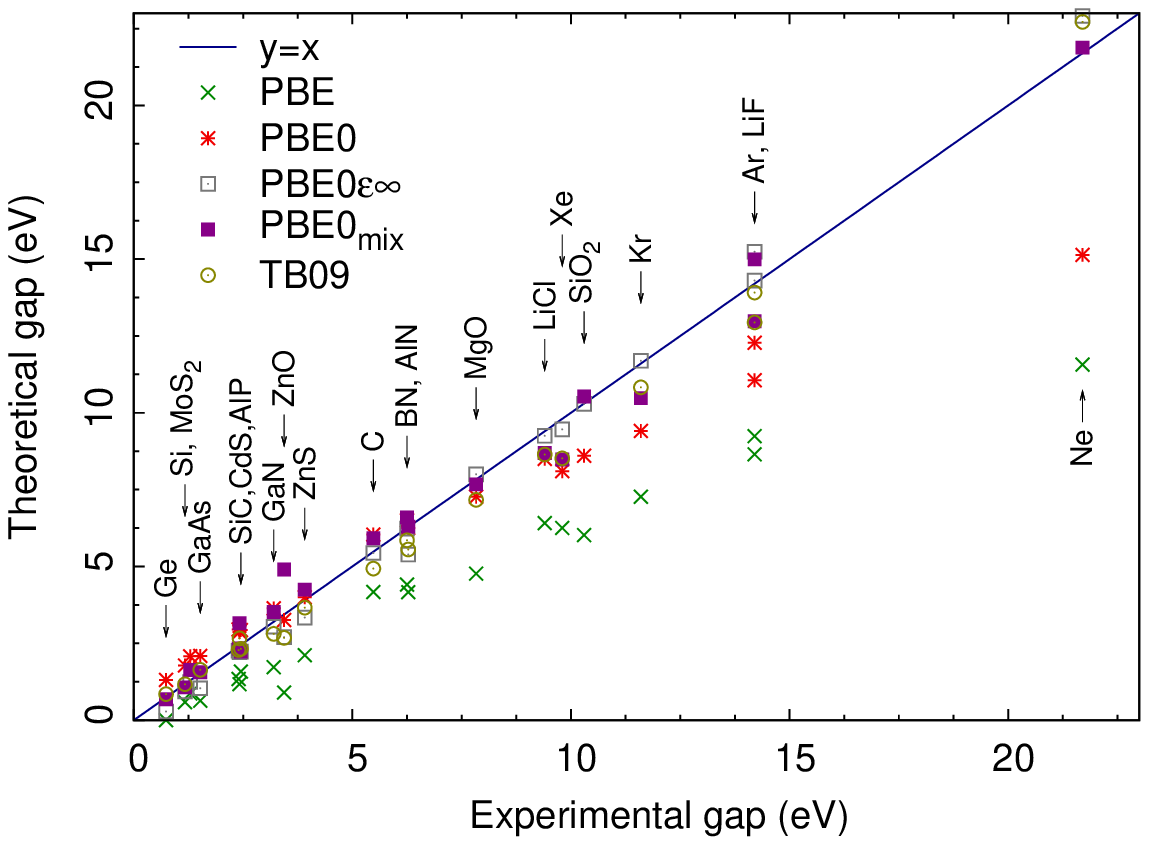} 
 \includegraphics[width=.45\textwidth]{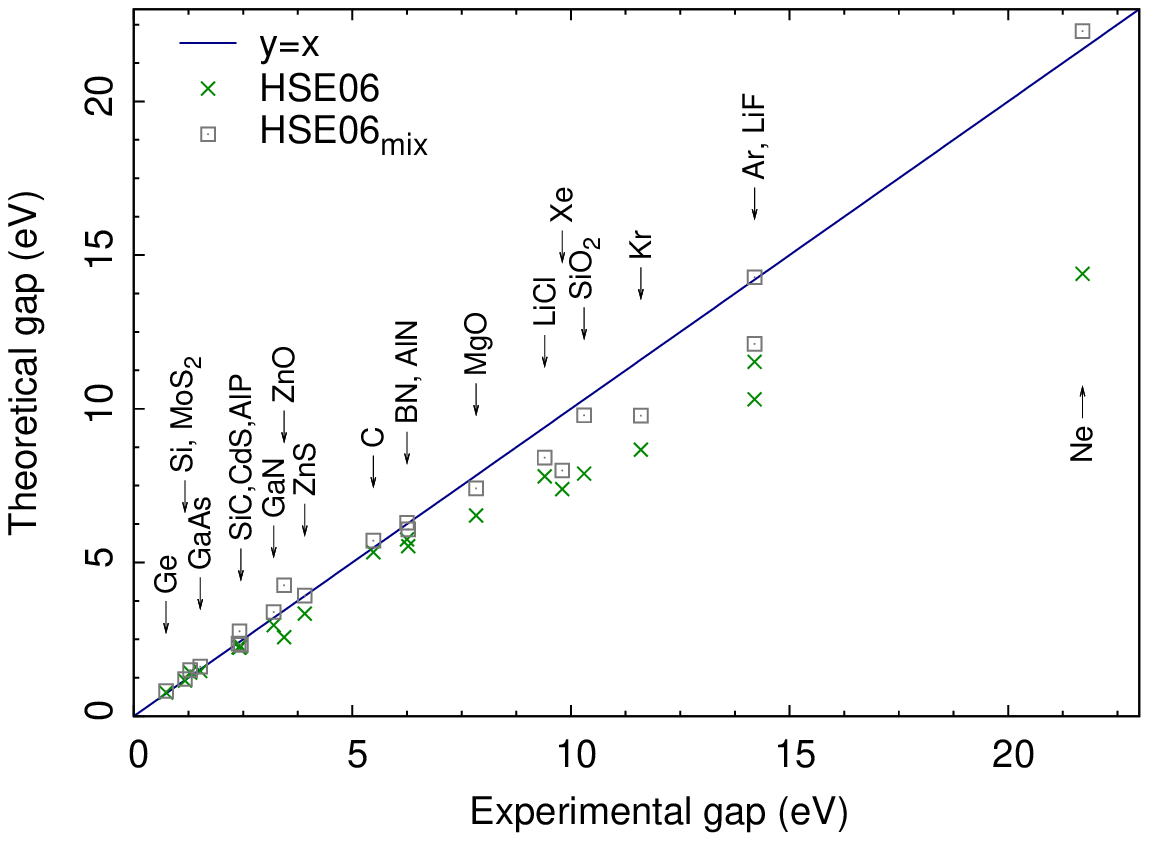} 
 \caption{(Color online) Electronic band-gaps calculated as
    differences of generalized KS eigenvalues for a series of
    semiconductors. All calculations were performed at the
    experimental lattice constant. The symbols labeled
    PBE0$_{\epsilon_\infty}$, PBE0$_\text{mix}$ and HSE06$_\text{mix}$ are
    the results obtained with the hybrid functionals proposed in this
    Article. TB09 results come from Ref.~\onlinecite{tran09}.
 \label{fig:gaps}
 }
\end{center}
\end{figure*}

In Fig.~\ref{fig:gaps} we compare the band-gaps calculated using
$\alpha=1/\epsilon^\text{PBE}_\infty$ as the mixing parameter of a
modified PBE0 hybrid functional (red stars labeled
PBE0${\epsilon_\infty}$), compared to experimental data and other
theoretical results. A more detailed comparison together with average
errors can be found in the table given as supplemental materials.  We
can see that the hybrid PBE0 ($\alpha=1/4$)~\cite{ernzerhof99}
already improves dramatically the result with respect to a local PBE
functional, bringing the calculated gaps towards the experimental
values with an average error of less than a 30\%. However, this number
alone hides the fact that PBE0 gives excellent results for some
intermediate band gap materials, like diamond, BN, AlN, etc., whereas
it fails both for small band gap materials (like Si, or Ge),
overestimating their gaps, and for large band gap materials (like the
rare-gases) where the gaps are underestimated. This is not surprising:
In materials like Si, electrons are delocalized and easily
polarizable, leading to a strong screening and small mixing (the
optimal mixing is actually $\alpha_\text{opt}=0.12$). For Ne,
electrons are localized, and screening is basically nonexistent
($\alpha_\text{opt}=0.70$).  This effect is captured by the dependence
on the dielectric constant of the mixing with $\alpha =
1/\epsilon_{\infty}$.  Such a calculation decreases overall the error
by almost a factor of two. The remaining error is dominated by the
large underestimation of the gap for materials like Si, Ge, or GaAs.
This point to the fact that, in order to predict good gaps approaching
the metallic limit, one needs a finite amount of Fock exchange that is
not accounted for by the simple $1/ \epsilon_\infty$ model for the
mixing, as it can also be seen in Fig.~\ref{fig:epsvsmix}.

The ${\epsilon_\infty}$-dependent mixing is hence physically motivated
and can yield good band gaps, however, the procedure to obtain it
(i.e. the calculation of the dielectric constant) is fairly cumbersome
and therefore often unpractical.  The best option would be to find an
estimator for the dielectric constant from quantities readily
available from the ground-state.  To obtain such relation, we first
observe that there is a strong correlation between the electronic
dielectric constant of the material and the energy
gap~\cite{fiorentini95}. Of course, it is not desirable to have a
functional depending explicitly on the band-gap of the material, so
the question is if one can find an estimator of this quantity based
solely on reduced densities.

In fact, several density-functional estimators of a ``local gap'' have
been proposed in the past years. For example, Gutle et
al.~\cite{gutle99} proposed to use the quantity $G=\frac{1}{8}|\nabla
n|^2 / n^2$ to define the gap locally. Their arguments for the use of
this quantity were based on the asymptotic expansion of the function
$G$ for finite systems ($G$ will reduce to the ionization energy) and
on the observed piecewise exponential behavior of the
density~\cite{sperber71}. More recently, the quantity $|\nabla n|/n$
was used to model a position-dependent screening function in a
so-called ``local-hybrid'' functional\cite{heyd03,krukau08}. Also the
von Weizs\"{a}cker kinetic energy density $\tau_\text{W}=|\nabla
n|^2/8 n$ has been used to define a ``local band-gap'', and inserted
into a ``local-hybrid'' functional that turns off the exact-exchange
term when this local gap has metallic character~\cite{jaramillo03}.

To obtain a global estimator of the band-gap of the material, and
therefore of its static dielectric constant, we can average the local
estimator over the Brillouin zone. We will follow the idea contained
in the meta-GGA of Tran and Blaha (TB09)~\cite{tran09} and define the
quantity
\begin{equation}
  \label{eq:barg}
  \bar g = \frac{1}{V_\text{cell}} \int_\text{cell}\!\!\D^3{r}\; 
  \sqrt\frac{|\nabla n(\br)|}{n(\br)}
  \,,
\end{equation}
where the integral is over the unit cell of volume $V_\text{cell}$. We
note that the quantity $\bar g$ is very similar to the average involved
in the calculation of the $c_\text{TB09}$ parameter of TB09,
and is quite stable regardless on which theory is used to evaluate the density. 
In fact, we verified that using as input either a PBE or a Hartree-Fock
density leads to only minor differences in its value. Our basic
hypothesis is that the mixing parameter can be written as a function
of the parameter $\bar g$.

To proceed we need to specify the local part of the hybrid.  We chose
to use PBE0 form~\cite{ernzerhof99} which is the basis for the
screened hybrid Heyd, Scuseria and Ernzerhof (HSE)~\cite{heyd03}.
As we expected, there is a clear correlation
between the value of $\alpha_\text{opt}$ and $\bar g$; it can be quite
well fitted by the simple function $\alpha(\bar g)$
\begin{equation}
  \label{eq:fitpbe0}
  \alpha = -1.00778 + 1.10507 \; \bar g
  \,,
\end{equation}
with $\bar g$ in atomic units.

Analyzing the resulting gaps displayed in Fig.~\ref{fig:gaps} (labeled
PBE0$_\text{mix}$) we realize that
fixing the mixing parameter according to
Eq.~\ref{eq:fitpbe0}~\cite{note2} reduces the mean average error to
slightly more than 14\%, much better than PBE0, and slightly better
than the HSE06~\cite{krukau06} screened hybrid functional.
Furthermore, in contrast with PBE0 and HSE06, our density-dependent
mixing describes equally well small, medium and large gap systems.
The largest errors arise for $d$-electron materials like ZnO where our
recipe overestimates the mixing parameter and therefore the electronic
gap. This is due to the fact that localized $d$-states give too large
contributions to the mixing through strong density variations. A
possible solution would be finding a more pertinent density-estimator
for those $d$ states.

In view of the success of screened hybrids in improving the accuracy
of PBE0, we applied our construction also to the HSE06
functional~\cite{krukau06}. In this case, the physical interpretation
of the mixing parameter as an inverse screening is considerably more
complicated, as screening is already present to some extent in the
range separation. Following our protocol, we arrived at the following
fit for the mixing parameter
\begin{equation}
  \alpha = 0.121983 + 0.130711\; \bar g^4
  \,.
\end{equation}
We remark the different power of $\bar g$ in the expression. This is
due to the fact that the screening already present in HSE06 decreases
considerably the strength of the Hartree-Fock term, increasing the
values of $\alpha_\text{opt}$ required to reproduce the experimental
band gap of small-gap systems. This is actually the cause of the
success of HSE06: for example, for Si $\alpha_\text{opt}$ is now 0.24,
which is very close to the actual mixing of HSE06 ($\alpha = 1/4$).
Results for band-gaps using HSE06 and our new mixing scheme (labeled
HSE06$_\text{mix}$) are shown in Fig.~\ref{fig:gaps}). Our new mixing
scheme brings down the HSE06 error from around 17\% to 10\%, of the
same order of magnitude as the error incurred by the $G_0W_0$
approximation and the new TB09 meta-GGA~\cite{tran09}.

We want now to compare our approach to the the new TB09 meta-GGA. The
physical interpretation of our functional implies that the mixing
parameter should take values between 0 and 1, while in TB09 the
corresponding parameter $c_\text{TB09}$ is always larger than one.
This difference stems from the fact that the TB09 functional is a
purely local potential.  Therefore, the band-gap defined in terms of
total energy differences should be equal to the difference of the KS
eigenvalues plus the derivative discontinuity of the xc potential.  It
is known that the Becke-Johnson potential~\cite{becke06} (upon which
the TB09 functional has been constructed) reproduces to a very good
extent the derivative discontinuity of exact
exchange for molecular chains~\cite{armiento08}.  Therefore, it is reasonable to expect
that the DFT band-gap with the TB09 functional, after adding the
derivative discontinuity, would actually become much larger than
experiment, and possibly even larger than the Hartree-Fock gap since
$c_\text{TB09}$ is always larger than one, in agreement with what was
proved by Gr\"uning \etal~\cite{gruning06}.  Of course, the aim of
TB09 is to obtain the gaps simply as eigenvalue differences, and
therefore require $c_\text{TB09} > 1$ in order to compensate for the
underestimation of the eigenvalue gap in the Becke-Johnson
functional~\cite{tran07}. In this sense, it is similar in spirit to
the pragmatic x$\alpha$ approach~\cite{xalpha}.

From Fig.~\ref{fig:gaps} it is clear that the errors of TB09 is
comparable to ours. On the one hand, the meta-GGA is clearly much
lighter from the computational point of view than any hybrid
functional. On the other hand, TB09 is an approximation for the xc
{\it potential}, and is thus incapable of yielding total energies (in
contrast to our approach). In fact, it can be proved that TB09 is not
the functional derivative of any energy functional, and therefore
violates serious constraints, like the zero-force
theorem~\cite{gaiduk09}. As a consequence, such functionals do not
allow to calculate structural properties.  Our approach, instead, can
also be used to calculate total energies and structural properties.
We tested our functionals and found that they give relaxed geometries
as good as the standard PBE0
and HSE06 (with lattice constants better than 0.7\% for the cubic semiconductors
considered here).  Moreover, we mention that the TB09 meta-GGA also
inherited some of the problems of the Becke-Johnson
functional~\cite{becke06} on which it is based (for example, TB09 is
not size-consistent and gauge invariant~\cite{rasanen10}).

It is clear that the averaging procedure in Eq.~\eqref{eq:barg} is
only meaningful for bulk systems and will fail for finite systems,
slabs, interfaces, etc.  This issue can be fixed by converting the
global $\bar g$ (and thus the uniform screening) into a local function
function $\bar g({\bf r})$, i.e. transforming the functional in a
local hybrid. The function $\bar g({\bf r})$ can be determined by
restricting the integral of Eq.~\eqref{eq:barg} to a neighborhood of
${\bf r}$, with a range related to the screening length of the system.
This screening length can be taken as constant, as in the HSE functionals,
or can be a local functional of the density (but clearly not a function of the average
density). This procedure would not change significantly the results for the systems studied
here as they all have very small unit cells.

In conclusion, we proposed a scheme to calculate on-the-fly the mixing
parameter in hybrid functionals depending on the density of the
system. In this way, the average error
on the values of the energy gap were considerably reduced with
respect to the original hybrid functionals. 
The resulting band-gaps are roughly of the same
quality as those obtained using a \GW\ approach 
or the new meta-GGA of Tran and Blaha. Moreover,
this method assures also an excellent description of the structural
properties. These improvements 
are obtained with no increase of computational
time with respect to a fixed mixing parameter.

Part of the calculations were performed at the LCA of the University
of Coimbra and at GENCI (project x2010096017). We acknowledge funding
from the European Commission through the e-I3 ETSF project (Contract
\#211956). MALM, JV, and SB acknowledge partial funding from the
French ANR (ANR-08-CEXC8-008-01), and from the program PIR Mat\'eriaux
-- MaProSu of CNRS.

%\bibliography{biblio}

\begin{thebibliography}{99}
\bibitem{note} Note that this is similar to the sX LDA method~\cite{bylander90,seidl96},
where $\Sigma$ is replaced by a Thomas-Fermi-screened exchange part and
the LDA xc functional.
\bibitem{note2} Note that in principle there is a contribution to
the potential due to the variation of the mixing parameter $\alpha$
with the density. In these calculations we have neglected this
term. However, we believe this to be a very good approximation due to
the robustness of $\bar g$ to the actual details of the calculation of the density.
\end{thebibliography}

\end{document}

\title{Supplemental material to the paper ``Density-based mixing parameter for hybrid functionals''}

\author{Miguel A. L. Marques}
\affiliation{LPMCN, Universit\'e Claude Bernard Lyon I and CNRS, 69622
Villeurbanne, France}
\affiliation{European Theoretical Spectroscopy Facility (ETSF)}

\author{Julien Vidal}
\affiliation{Institute for Research and Development of Photovoltaic Energy
(IRDEP), UMR 7174 CNRS/EDF/ENSCP, 6 quai Watier, 78401 Chatou, France}
\affiliation{Laboratoire des Solides Irradi\'es, \'Ecole Polytechnique, CNRS,
CEA-DSM, 91128 Palaiseau, France}
\affiliation{European Theoretical Spectroscopy Facility (ETSF)}

\author{Micael Oliveira}
\affiliation{LPMCN, Universit\'e Claude Bernard Lyon I and CNRS, 69622
Villeurbanne, France}
\affiliation{European Theoretical Spectroscopy Facility (ETSF)}

\author{Lucia Reining}
\affiliation{Laboratoire des Solides Irradi\'es, \'Ecole Polytechnique, CNRS,
CEA-DSM, 91128 Palaiseau, France}
\affiliation{European Theoretical Spectroscopy Facility (ETSF)}

\author{Silvana Botti}
\affiliation{Laboratoire des Solides Irradi\'es, \'Ecole Polytechnique, CNRS,
CEA-DSM, 91128 Palaiseau, France}
\affiliation{LPMCN, Universit\'e Claude Bernard Lyon I and CNRS, 69622
Villeurbanne, France}
\affiliation{European Theoretical Spectroscopy Facility (ETSF)}
% \date{\today}

\begin{table*}[t]
\begin{center}
  \caption{\label{tab:gaps} Electronic band-gaps calculated as
    differences of Kohn-Sham eigenvalues for a series of
    semiconductors. All calculations were performed at the
    experimental lattice constant. The column label HF+c denotes
    Hartree-Fock including PBE correlation. The columns
    PBE0$_{\epsilon_\infty}$, PBE0$_\text{mix}$ and HSE06$_\text{mix}$ present
    the results obtained with the hybrid functionals proposed in this
    Article. The columns TB09 and $G_0W_0$ are from [F. Tran and P. Blaha, Phys. Rev. Lett. {\bf 102},
  226401 (2009)].}

\begin{tabular}[c]{ldddddddddd}
  & \multicolumn{1}{c}{exp.}
  & \multicolumn{1}{c}{PBE}
  & \multicolumn{1}{c}{HF+c}
  & \multicolumn{1}{c}{PBE0}
  & \multicolumn{1}{c}{PBE0${\epsilon_\infty}$}
  & \multicolumn{1}{c}{PBE0$_\text{mix}$}
  & \multicolumn{1}{c}{HSE06}
  & \multicolumn{1}{c}{HSE06$_\text{mix}$}
  & \multicolumn{1}{c}{TB09}
  & \multicolumn{1}{c}{$G_0W_0$} \\
\hline\\[-0.15cm]
Ne     & 21.70 & 11.57 & 26.14 & 15.14 & 22.95 & 21.88 & 14.39 & 22.29 & 22.72 &
19.59 \\
Ar     & 14.20 &  8.65 & 18.45 & 11.06 & 14.35 & 12.98 & 10.31 & 12.11 & 13.91 &
13.28 \\
Kr     & 11.60 &  7.27 & 16.04 &  9.41 & 11.75 & 10.48 &  8.67 &  9.78 & 10.83 &
      \\
Xe     &  9.80 &  6.25 & 13.79 &  8.10 &  9.53 &  8.48 &  7.39 &  7.99 &  8.52 &
      \\
C      &  5.48 &  4.17 & 12.05 &  6.06 &  5.48 &  5.92 &  5.33 &  5.71 &  4.93 &
 5.50 \\
Si     &  1.17 &  0.59 &  6.00 &  1.78 &  0.98 &  1.07 &  1.16 &  1.21 &  1.17 &
 1.12 \\
Ge     &  0.74 &  0.00 &  5.49 &  1.31 &  0.32 &  0.68 &  0.77 &  0.82 &  0.85 &
 0.66 \\
LiF    & 14.20 &  9.24 & 21.55 & 12.26 & 15.27 & 14.99 & 11.53 & 14.28 & 12.94 &
13.27 \\
LiCl   &  9.40 &  6.41 & 14.94 &  8.50 &  9.28 &  8.69 &  7.80 &  8.41 &  8.64 &
      \\
MgO    &  7.83 &  4.77 & 15.24 &  7.27 &  8.06 &  7.67 &  6.53 &  7.41 &  7.17 &
 7.25 \\
SiC    &  2.40 &  1.34 &  8.18 &  2.95 &  2.28 &  2.33 &  2.24 &  2.36 &  2.28 &
 2.27 \\
BN     &  6.25 &  4.41 & 13.06 &  6.50 &  6.25 &  6.60 &  5.75 &  6.29 &  5.85 &
 6.10 \\
GaN    &  3.20 &  1.72 & 10.29 &  3.64 &  3.07 &  3.52 &  2.96 &  3.39 &  2.81 &
 2.80 \\
GaAs   &  1.52 &  0.63 &  6.81 &  2.09 &  1.04 &  1.56 &  1.47 &  1.61 &  1.64 &
 1.30 \\
AlP    &  2.45 &  1.58 &  7.40 &  2.93 &  2.24 &  2.23 &  2.27 &  2.32 &  2.32 &
 2.44 \\
ZnS    &  3.91 &  2.11 & 10.06 &  4.00 &  3.38 &  4.25 &  3.34 &  3.92 &  3.66 &
 3.29 \\
CdS    &  2.42 &  1.17 &  8.56 &  2.87 &  2.25 &  3.15 &  2.23 &  2.76 &  2.66 &
 2.06 \\
AlN    &  6.28 &  4.16 & 12.94 &  6.25 &  5.39 &  6.29 &  5.53 &  6.08 &  5.55 &
 5.83 \\
SiO$_2$& 10.30 &  6.02 & 16.75 &  8.63 &  9.10 & 10.53 &  7.89 &  9.79 &       &
      \\
MoS$_2$&  1.29 &  0.87 &  7.90 &  2.09 &  1.25 &  1.63 &  1.42 &  1.50 &       &
      \\
ZnO    &  3.44 &  0.90 & 11.21 &  3.26 &  2.74 &  4.90 &  2.57 &  4.26 &  2.68 &
 2.51 \\
\hline\\[-0.15cm]
$\Delta$ (\%) & \multicolumn{1}{c}{--}
               & 47.32 &250.23 & 29.42 & 16.53 & 14.37 & 16.92 & 10.36 &  9.85 &
11.25 \\
\end{tabular}
\end{center}
\end{table*}